\documentclass{article}

\usepackage[square,sort,comma,numbers]{natbib}
\usepackage[final]{neurips_2021}

\usepackage[utf8]{inputenc} 
\usepackage[T1]{fontenc}    
\usepackage{hyperref}       
\usepackage{url}            
\usepackage{booktabs}       
\usepackage{amsfonts}       
\usepackage{nicefrac}       
\usepackage{microtype}      
\usepackage{array}
\usepackage{tabularx}
\usepackage{graphicx}
\usepackage{booktabs}
\usepackage{caption}
\usepackage{color}
\usepackage{xurl}
\usepackage{verbatim}
\usepackage{multirow}
\usepackage{xspace}
\usepackage{graphicx}
\usepackage{amsmath} 
\usepackage{bbm}
\usepackage{amssymb}
\usepackage{booktabs} 
\usepackage{times} 
\usepackage{lscape}
\usepackage{enumitem}
\usepackage[table,xcdraw]{xcolor}
\usepackage[normalem]{ulem}
\usepackage{xcolor}
\usepackage[innercaption]{sidecap}
\usepackage{lineno}
\usepackage{enumitem}
\usepackage{xcolor}
\usepackage{caption}
\usepackage{dcolumn}
\usepackage{subfigure}
\usepackage{bbm}

\newcommand{\modelname}{\textsc{MashNet}\xspace}
\newcommand{\std}[1]{\scriptsize{$\pm$#1}}

\title{Deep Learning for Reference-Free Geolocation of Poplar Trees}

\author{
  Cai W. John\thanks{Shared first authorship} \\
  Bredesen Center\\
  University of Tennessee, Knoxville\\
  Knoxville, TN 37996 \\
  \texttt{cjohn3@vols.utk.edu} \\
   \And
  Owen Queen$^*$\thanks{Work done while at University of Tennessee, Knoxville}\\
  Department of Biomedical Informatics\\
  Harvard Medical School\\
  Boston, MA 02115 \\
  \texttt{owen\_queen@hms.harvard.edu} \\
   \And
    Wellington Muchero\\
    Center for Bioenergy Innovation\\
    Oak Ridge National Laboratory\\
    Oak Ridge, TN 37830 \\
    \texttt{mucherow@ornl.gov}
   \And
   Scott J. Emrich \\
   Electrical Eng. and Computer Science\\ 
   Bredesen Center\\
   University of Tennessee, Knoxville\\
   Knoxville, TN 37996 \\
   \texttt{semrich@utk.edu}
}

\begin{document}

\maketitle

\begin{abstract}
    A core task in precision agriculture is the identification of climatic and ecological conditions that are advantageous for a given crop. The most succinct approach is geolocation, which is concerned with locating the native region of a given sample based on its genetic makeup. Here, we investigate genomic geolocation of \textit{Populus trichocarpa}, or poplar, which has been identified by the US Department of Energy as a fast-rotation biofuel crop to be harvested nationwide. In particular, we approach geolocation from a reference-free perspective, circumventing the need for compute-intensive processes such as variant calling and alignment. Our model, \modelname, predicts latitude and longitude for poplar trees from randomly-sampled, unaligned sequence fragments. We show that our model performs comparably to Locator, a state-of-the-art method based on aligned whole-genome sequence data. \modelname achieves an error of 34.0 km$^2$ compared to Locator's 22.1 km$^2$. \modelname allows growers to quickly and efficiently identify natural varieties that will be most productive in their growth environment based on genotype. This paper explores geolocation for precision agriculture while providing a framework and data source for further development by the machine learning community.

\end{abstract}

\section{Introduction}
    
Pollen dispersal in natural populations of \textit{Populus trichocarpa}, as well as other species, results in correlations between geography and genetic variation. These correlations can be leveraged to predict geographic origin of a sample from genetic data as demonstrated in previous studies \cite{locator} \cite{other_location_pred}. To date, all studies have achieved this prediction task using aligned, whole-genome sequence data. Here, we demonstrate our novel tool \modelname that predicts geographic origin from unaligned sequence fragments. We compare it to the current state of the art implementation, Locator \cite{locator}, which uses a deep learning architecture on aligned sequences. Our method performs similarly despite using more noisy sequence read-only information.

Sequence alignment is a necessary procedure to transform short read fragments into genome-scale information. Modern technology is only capable of sequencing small sections of DNA, so large-scale genotyping of individuals using sequencing data requires \textit{post hoc} alignment and variant-calling algorithms, usually relative to a well-established reference genome sequence \cite{Zhang18}. These algorithms are computationally intensive procedures that create major bottlenecks between sample collection and downstream analysis of variant data. Further, although reference genomes are increasingly common due to advances in both technology and assembly algorithms\cite{Nurk2020}, they still require large amounts of sequence data and resource intensive  \textit{de novo} assembly. These demands prevent many non-model organisms from being sequenced. Our approach is alignment-free and therefore can be applied to the many non-model organisms currently without a reference genome. It also circumvents the need for variant-calling algorithms allowing researchers to more rapidly analyze samples. For example, one can envision sampling natural genetic diversity in a species, and then using computational methods to suggest the ancestral origin(s) of unknown samples. This process is called geolocation. A simple spatial-climate map, such as the K\"{o}ppen-Geiger climate system \cite{koppen}, could then map origin locations to desired growing environments. Being able to pinpoint these environments is key to precision agriculture.

In this study, we focus on \textit{Populus trichocarpa} (poplar) because of interest from the Department of Energy (DOE) in developing it as a fast-rotation biofuel crop to be viable nationwide \cite{DOE}. Poplar's species range extends from southern California all the way to British Columbia encompassing a latitudinal range of 38.88 to 54.25 degrees \cite{Slavov_2012}. This range includes a diversity of macro and micro-environments that have likely shaped subpopulations of this species. Our goal is to predict the latitudinal and longitudinal coordinates of these genotypes from their sequence data, a task known as genomic geolocation.

Geolocation has applications in precision agriculture. When considering a new site for a tree nursery it is desirable to clone samples well-suited to that environment. Given that these trees have often been previously cloned, and relocated to common gardens and greenhouses for commercial use and agricultural research, it can be difficult to obtain meta-data locating them to their origin environment. \modelname resolves this issue allowing growers to rapidly identify the origin location of their trees, and identify which will be most productive in the new climate.

In this work, we present \modelname, a deep learning-based model that can perform accurate geolocation of poplar trees. The model uses a multi-task neural architecture to jointly predict latitude and longitude coordinates for each sample. Importantly, this method uses Mash sketches \cite{mash}, an alignment-free feature extraction method that randomly samples k-mers from sequencing read data. We demonstrate that \modelname can use alignment-free Mash sketches to compete with WGS-based methods. We open source our methods and data while highlighting the importance of this task to precision agriculture.

\section{Methods}

\subsection{Data}
\begin{figure}[t!]
    \centering
    \scalebox{0.65}{
    \includegraphics{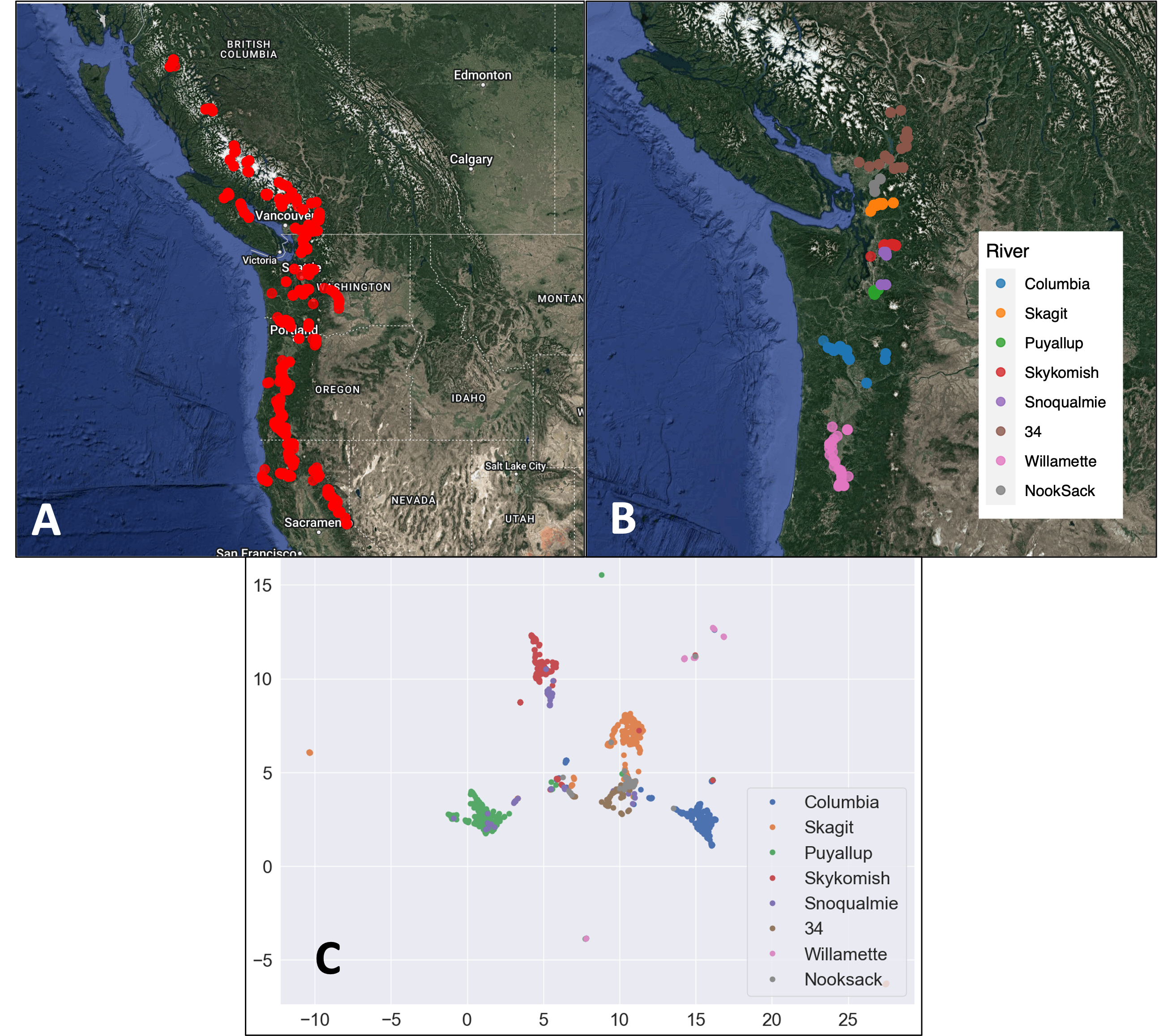}
    }
    \caption{A) Map of the origin location of all 1,252 poplar samples. B) Reduced set of 919 samples used for PCA-UMAP clustering by river system. Downsampling from 1,252 samples is achieved by retaining only river systems with $\ge$ 35 members. C) PCA-UMAP embedding of 919 clustering samples colored by river system.}
    \label{fig:maps}
\end{figure}

We consider 1,252 poplar genotypes from a representative sampling of the latitudinal distribution of its species range (see Figure \ref{fig:maps} panel A). Genome re-sequencing, alignment and variant-calling of this population was previously described by Zhang et al. \cite{Zhang18}. We use these aligned and variant called sequences in Locator as a performance benchmark for our alignment-free method. \modelname is trained on unaligned reads. Out of the total 1,252 samples, 1,024 have reads that are publicly available for download from the NCBI's Sequence Read Archive (SRA). A map of sample ID's to SRA key is included with the meta data in our Github \footnote{All codes and data found at https://github.com/owencqueen/MashPredict}. During training, meta data labels with ground-truth latitude and longitude coordinates for all 1,252 samples are used. These are also included on our GitHub repository.  Unfortunately, we are unable to publicly host the aligned WGS used to train and test Locator, as well as the remaining 228 sample reads. This is due to current access restrictions.

Associated with each sample are several meta-data variables. The first is river system, which corresponds to the nearby river from which each sample was originally collected. This variable in particular shows strong signal, as is evidenced by Figure \ref{fig:maps}C, which shows a PCA-UMAP \cite{PCA-UMAP} projection of each sample colored by its associated river system. This projection illustrates the correlation between origin location and genotype that we will leverage to geolocate these samples.

\subsection{MinHashing Unaligned Reads}
A major innovation of this work is achieving prediction from unaligned reads. We accomplished this using the Mash software \cite{mash}. This process uses read fragments to create a reduced representation of the genome, \textit{i.e.}, a ``sketch'' of the genome, which has been shown to accurately reflect genome-wide structure \cite{mash}. It does this by randomly sub-sampling k-mer's from the read fragments using a minHash-based approach. When using Mash the user must define the k-mer length to use, as well as the number of hash functions to store which determines the sketch size ($s$). For our study, we chose a k-mer length of 21. This is the default in Mash and their studies demonstrate this k-mer length robustly maps to Average Nucleotide Identity (an alignment-based measure of mutation distance) across different sketch sizes.

Mash states that ``Increasing sketch size improves the accuracy of Mash estimates, especially for divergent genomes'' \cite{mash}.  To test this, we ran MASH at four different sketch sizes: $s$=500, 2000, 4000, and 50,000. We trained and tested our prediction algorithms across all four sketch sizes to compare performance (see Table \ref{tab:num_v_performance} and Figure \ref{fig:error}).

Once sketched, we devised a novel application of the Mash output. The input to Mash is a dataset of $n$ samples of reads $R_i$ that correspond to sequencing reads for a given poplar tree, $\mathcal{D} = \{R_1, ..., R_n\}$. Assuming no hash collisions, each hash function $H_i$ is a unique identifier for a 21 length k-mer. Mash samples $s$ random k-mers per $R_i$, thereby resulting in a set of $s$ hash functions, known as a sketch: $M_i = \{ H_1^i, ..., H_s^i\}$. $s$ is a user-defined parameter called sketch size that is discussed in subsequent sections. This procedure is repeated for every sample in $\mathcal{D}$ to build a set of sketches $\{M_1, ..., M_n\}$. Now, a union is taken over all hash functions in each sketch in order to construct a set of hash functions $\mathcal{H} = \bigcup_{i=1}^{n} M_i$. Note that $|\mathcal{H}|$ is guaranteed to be upper-bounded by $s \times n$, but often $|\mathcal{H}| \ll s \times n$ because there are common k-mers shared across samples $R_i$.

Typically, these sketches are used for a simple pairwise comparison of genomes to estimate genetic distance. For a pair of genotypes, this is done by set comparison of the hash functions in each genome sketch, such as a Jaccard index. Here, instead of only looking at pairwise comparisons, we look at set membership across the entire population. This is achieved by building a presence-absence matrix for the hash functions in each sketch. Taking the set of all hash functions $\mathcal{H}$, we construct a vector by placing a 1 if the hash is found in sketch $M_i$ and a 0 if it is not found in sketch $M_i$. Formally, each vector representation $V_i$ corresponding to a sketch $M_i$ is defined by $V_i =  \{ \mathbbm{1}_{[H_j \in M_i]} | H_j \in \mathcal{H} \}$ where the indicator function $\mathbbm{1}$ sets the value to 1 if $H_j \in M_i$ and 0 otherwise. This converts each set $M_i$ to a constant-size binary vector $V_i$. Assuming no hash collisions, this means our matrix represents a random sampling of k-mers, with a 1 indicating that k-mer as present in a genotype and 0 indicating its absence. This provides a binary input matrix for our deep learning architecture \modelname.

\subsection{\modelname Model}

\modelname is a neural network for prediction and representation of Mash sketches. This network takes the binary Mash matrix as input and performs predictions for latitude and longitude. The model architecture consists of a combination of linear and LayerNorm \cite{ba2016layer} layers followed by ELU \cite{clevert2015fast} activation functions. We also chose to use a Batch Normalization \cite{ioffe2015batch} layer to process the input, following Locator's \cite{locator} similar decision. We empirically found that this architecture improved performance on the sparse Mash sketch input (see Figure \ref{fig:error}). 

\modelname can be used for prediction of any phenotype, but we chose to focus it on geolocation, \textit{i.e.}, predicting latitude and longitude coordinates for each sample. As the output of the network, we have a multi-task learning setup, where we jointly predict both latitude and longitude in the same forward pass. The \modelname model $F$ takes a vectorized Mash sketch $V_i$ as input and outputs a coordinate $\mathbb{R}^2$. Our loss function is a simple Absolute Error (AE) with equal weight for both latitude and longitude, \textit{i.e.}, $L = L_{\text{lat}} + L_{\text{long}}$, where $L_{\text{lat}}$ is the AE for latitude and $L_{\text{long}}$ is the AE for longitude. 

\subsection{Experiments and Comparison Models}

For geolocation, we compare \modelname to several other non-neural models. First, we use $k$-nearest neighbors (kNN) on the Mash distances. Mash computes pairwise distances with a set-based distance function that approximates the Jaccard index between each sample, as discussed in \cite{mash}. We compute this pairwise distance matrix and use this as a distance metric in the kNN prediction. Additionally, XGBoost and ElasticNet algorithms are employed on the binarized Mash sketches. For each model, we perform a search over a hyperparameter space to optimize model performance: for kNN, we search over k values, for XGBoost and ElasticNet, we search over parameters controlling regularization strength and learning rate.

We also compare several WGS methods against models trained on sketch-based inputs. First, we use a state-of-the-art method Locator \cite{locator}, which was designed for direct geolocation prediction from WGS data. Finally, we use XGBoost \cite{chen2016xgboost} and ElasticNet \cite{zou2005regularization} algorithms on a principal component analysis (PCA)-reduced representation. PCA is used to reduce the WGS representations because of the large size and high level of sparsity. PCA is a widely established technique in bioinformatics, and it has previously shown to be effective in compressing WGS samples \cite{PCA-UMAP}.

Each experiment is performed with 30 separate 5-fold cross validations, each with individual random seeds. Performance metrics are averaged across all folds for one cross validation, and we report the mean and standard error across all 30 cross validations for each separate experiment. Each error in Table \ref{tab:num_v_performance} is reported as mean absolute error (MAE) in kilometers, which is calculated from latitude and longitude coordinates via geodesic distance provided by the geopy package \cite{geopy}. We only use 5 trials of cross validation on Locator because of prohibitively long runtimes. For \modelname, we standard  scale the latitude and longitude before training and inverse scale the outputs to compute errors. This standard scaling approach involves transforming the data to a normal distribution with mean$=0$ and standard deviation$=1$. It seemed to have no detectable effect on performance for alternative models.

\section{Results}

\begin{figure}[h]
    \centering
    \scalebox{0.3}{
    \includegraphics{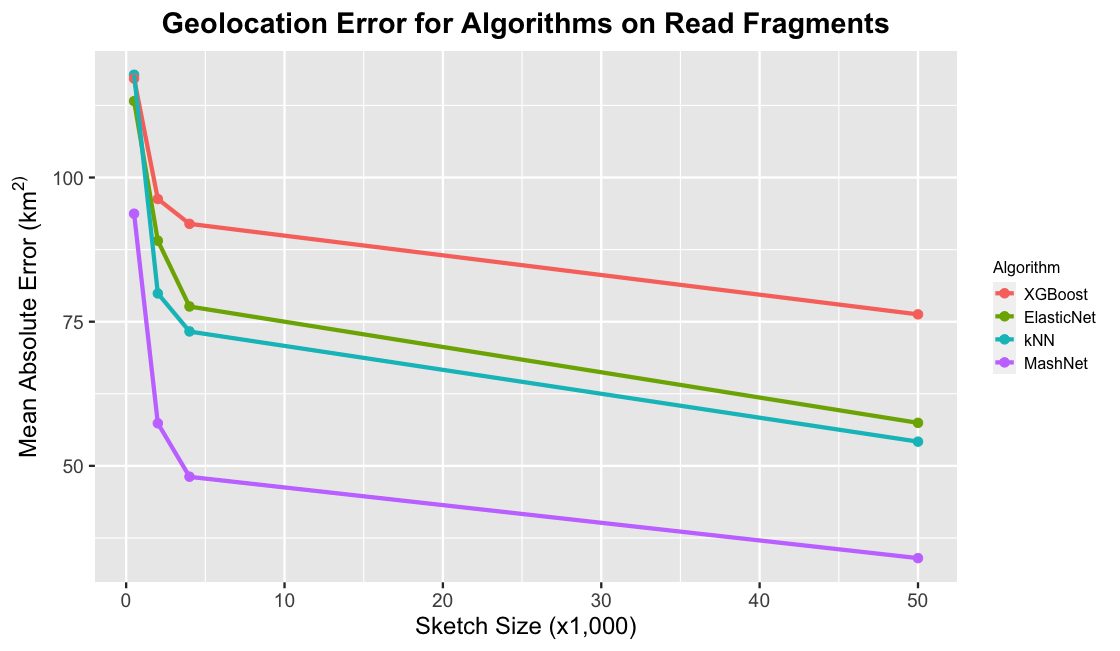}
    }
    \caption{Inspecting errors across varying sketch sizes for all algorithms applied to unaligned read fragments.}
    \label{fig:error}
\end{figure}

\begin{table}[h!]
    \centering
    \begin{subtable}
        \centering
        \begin{tabular}{c|c|c|c}
         & Locator & ElasticNet & XGBoost \\
         \midrule
        WGS & 22.10\std{1.37} & 236.54\std{0.02} & 37.77\std{0.09}
        \end{tabular}
        \vspace{3mm}
        \caption*{Table 1a}
    \end{subtable}
    \begin{subtable}
        \centering
        \begin{tabular}{c|c|c|c|c}
            Sketch size ($\times 10^3$) & kNN & ElasticNet & XGBoost & \modelname\\
            \midrule
            0.5 & 117.82\std{1.06} & 113.26\std{0.08} & 117.16\std{0.14} & 93.73\std{0.32} \\
            2 & 79.90\std{1.62} & 89.04\std{0.09} & 96.28\std{0.13} & 57.38\std{0.33} \\
            4 & 73.31\std{1.02} & 77.64\std{0.09} & 91.97\std{0.14} & 48.12\std{0.96} \\
            50 & 54.20\std{0.97} & 57.46\std{0.08} & 76.27\std{0.15} & 34.00\std{0.24} \\
        \end{tabular}
        \vspace{3mm}
        \caption*{Table 1b}
        \vspace{5mm}
    \end{subtable}
    \caption{Mean absolute error in kilometers$^2$ for various models trained on whole-genome sequence inputs (1a) and Mash sketch-encoded vectors (1b). \textit{Table 1a} ElasticNet and XGBoost are trained on PCA-reduced versions of SNP data obtained after sequence alignment. \textit{Table 1b} sketch size is shown in units of 1000 sketches. kNN is trained on Jaccard distance between each sample while all other methods are trained on vectorized Mash sketches.}
    \label{tab:num_v_performance}
\end{table}

Locator is the best-performing model, pinpointing the location to within 22.1km$^2$ of error. ElasticNet and XGBoost, which are both trained on PCA-reduced versions of the WGS SNPs, perform worse than Locator on the geolocation task. Within the Mash-based predictors, \modelname outperforms all methods, regardless of the sketch size. kNN performs better than both ElasticNet and XGBoost; this is likely because distance is defined based on the set-based metric used in the original Mash publication \cite{mash}. ElasticNet consistently outperforms XGBoost, with XGBoost being the least predictive model for Mash-based input data. 

Comparing across WGS and Mash-based predictors, WGS predictors perform better overall. This result is expected given the longer-range structure that is elucidated during the alignment procedure. However, several key patterns emerge. First, \modelname still outperforms both WGS-based ElasticNet and XGBoost when using a sketch size of 50,000. This highlights the utility and capacity of \modelname and neural networks for geolocation, even from noisy data such as Mash sketches. Second, on the WGS data XGBoost outperforms ElasticNet, but on the Mash-based input ElasticNet performs better. This is most likely due to the differences in data geometry. The Mash-based input data are sparse, binary vectors while PCA-reduced WGS inputs are dense with fewer dimensions. The geolocation task is highly nonlinear, so in the dense WGS setting, we expect a tree-based model (XGBoost) to perform better than a linear model (ElasticNet).

We also perform benchmarking across different numbers of Mash sketches. Sketch size is an important tuning factor when using \modelname. As seen in Table \ref{tab:num_v_performance}, performance increases with increasing sketch size. In Mash, compute time to build a sketch is largely invariant to sketch size, however overall computational costs will increase due to higher dimensional input being passed to downstream prediction models. This is a trade-off that must be managed. In general, traditional, non-deep learning-based methods (ElasticNet and XGBoost) perform poorly on Mash sketches, highlighting the need for an alternative such as our model \modelname. However, the set-based distance metric leveraged by the original Mash publication has been further validated here, showing a clear ability to recover significant predictive signal using kNN, which even outperforms more sophisticated methods such as ElasticNet and XGBoost.

\section{Discussion}

The genome sciences contain many applications for reference-free prediction using computational techniques. To the best knowledge of the authors, this study is one of the first attempts at trait prediction from unaligned read fragments. Innovations in this space have the potential for large impact on topics ranging from precision agriculture to medical diagnostic tools. 

In this study, we present a solution to the challenging task of geolocation of poplar trees from unaligned read fragments. We approach this problem by leveraging a commonly-used bioinformatics tool, Mash, and create a framework that can circumvent the computationally expensive procedures of genome assembly and short read alignment. Our solution, \modelname, uses a neural network to predict latitude and longitude coordinates for each sample, achieving within 12.1 km$^2$ prediction accuracy to the state-of-the-art whole-genome sequence-based method, Locator \cite{locator}.  

Future studies will attempt to improve our predictive capacity using unaligned reads. The initial studies undertaken in this paper outline two paths to improvement. The first is to try to pre-identify important k-mers on which screening should be focused. For example, in currently unpublished work we have identified regulatory hotspots through genome-wide association (GWAS) mapping of climatic variation. We hypothesize that if we could sample k-mer's directly from these hotspots--- and not randomly as we do currently--- we could focus on the higher variance regions and therefore significantly boost prediction performance. However, this approach would require \textit{a priori} knowledge of the genomic location of these hotspots and therefore pre-existing aligned WGS data. Thus, while such a hybrid approach would likely improve predictive performance, it would also nullify the generalizability of our \modelname approach to non-model organisms.

A second approach would be to increase the sketch size of the minHashing procedure. In Figure \ref{fig:error}, we observe that there seems to be a performance plateau associated with increasing sketch size. We hypothesize this occurs once sufficient sampling coverage of the genome has been achieved. This suggests that while increasing sketch size would lead to performance gains, these gains are likely to be marginal. This presents an open question: \modelname can predict locations within 34km$^2$, but could a more advanced technique predict these locations with less error?

Given the importance of the geolocation task for precision agriculture, we present this as an \textbf{open problem for the machine learning community}. Our tool, \modelname, demonstrates how deep learning can achieve impressive results on reference-free geolocation tasks, even when compared to state-of-the-art models based on WGS representations. We believe that more advanced tools can be developed for this area and used to improve prediction accuracy of the ideal ecosystem in which a crop should be grown. We open-source the codebase and datasets used for this study with the hope that future development will focus on new techniques for representing unaligned, fragmented reads for machine learning, as well as more sophisticated prediction architectures.

\bibliographystyle{unsrt}
\bibliography{citations.bib}

\end{document}